\def\showall{0}
\def\useieeelayout{0}

\let\ampb&
\if\useieeelayout1
\newcommand\arxivbreak{}
\else
\newcommand\arxivbreak{\\\ampb}
\fi
\if\useieeelayout1
\newcommand\ifacbreak{\\\ampb}
\else
\newcommand\ifacbreak{}
\fi

\newcommand{\myfrac}[2]{
\if\useieeelayout1
\tfrac{#1}{#2}
\else
\dfrac{#1}{#2}
\fi
}

\newcommand{\inACC}[1]{\if\useieeelayout1{#1}\fi\if\showall1{\color{green!50!black}In ACC: #1}\fi}

\newcommand{\inArxiv}[1]{\if\useieeelayout0{#1}\else\if\showall1{\color{blue}In ArXiV: #1}\fi\fi}

\if\useieeelayout1
\documentclass{ifacconf}
\else
\documentclass{article}
\usepackage{arxiv}
\fi

\usepackage{graphicx}
\usepackage{natbib}
\usepackage{amsmath}
\usepackage{amsfonts}
\usepackage{xcolor}
\usepackage{booktabs}
\usepackage{balance}
\usepackage{fancyvrb}
\if\useieeelayout0
\usepackage{amsthm}
\newtheorem{thm}{Theorem}
\newtheorem{defn}{Definition}
\newtheorem{lem}{Lemma}
\newtheorem{pf}{Proof}
\newtheorem{prop}{Proposition}
\newtheorem{rem}{Remark}
\usepackage{amssymb}
\usepackage{accents}
\newcommand{\ubar}[1]{\underaccent{\bar}{#1}}
\usepackage{hyperref}
\hypersetup{
    colorlinks = true,
    allcolors = black!50!blue
}

\fi
\newtheorem{problem}{Problem}
\newtheorem{assumption}{Assumption}

\newcommand{\R}{\mathbb{R}}

\newcommand{\diag}{\textup{diag}}

\usepackage[numbered,framed]{matlab-prettifier}
\begin{document}
\if\useieeelayout1
\begin{frontmatter}
\title{A Robust Observer with Gyroscopic Bias Correction for Rotational Dynamics}

\author[First]{Erjen Lefeber}
\author[Second,Third]{Marcus Greiff}
\author[Second]{Anders Robertsson}

\address[First]{Dept.\ of Mech.\ Engineering, Eindhoven University of Technology, Eindhoven, The Netherlands (e-mail: A.A.J.Lefeber@tue.nl).}
\address[Second]{Department of Automatic Control, Lund University, Lund, Sweden (e-mail: [Marcus.Greiff,Anders.Robertsson]@control.lth.se)}
\address[Third]{Mitsubishi Electric Research Laboratories, Cambridge, MA, U.S.A. (e-mail: Greiff@merl.com)}

\thanks[footnoteinfo]{This research was partly funded by the ELLIIT-project "Autonomous Radiation Mapping and Isotope Composition Identification
by Mobile Gamma Spectroscopy" and the SSF project “Semantic mapping and visual navigation for smart robots” (RIT15-0038).}
\else

\title{A Robust Observer with Gyroscopic Bias Correction for Rotational Dynamics}
\author{
\begin{tabular}[t]{ccccccccc}
        $\;\;$Erjen Lefeber$^1$ &&$\qquad$&&Marcus Greiff$^2$ &&$\quad$&&Anders Robertsson$^3$\\
        {\normalfont TU Eindhoven} &&$\qquad$&& {\normalfont MERL} &&$\quad$&& {\normalfont Lund University}
\end{tabular}
}
\footnotetext[1]{\emph{Dep. of Mech. Engineering, Eindhoven University of Technology, Eindhoven, The Netherlands,} \texttt{A.A.J.Lefeber@tue.nl}.}
\footnotetext[2]{\emph{Mitsubishi Electric Research Laboratories, Cambridge, MA, U.S.A.,} \texttt{Greiff@merl.com}.}
\footnotetext[3]{\emph{Department of Automatic Control, Lund University, Lund, Sweden,} \texttt{Anders.Robertsson@control.lth.se}.}
\footnotetext[4]{This research was partly funded by the ELLIIT-project "Autonomous Radiation Mapping and Isotope Composition Identification
by Mobile Gamma Spectroscopy" and the SSF project “Semantic mapping and visual navigation for smart robots”.}
\maketitle
\fi
\begin{abstract}
We propose an observer for rotational dynamics subject to directional and gyroscopic measurements, which simultaneously estimates the gyroscopic biases and attitude rates. We show uniform almost global asymptotic and local exponential stability of the resulting error dynamics, implying robustness against bounded disturbances. This robustness is quantified with respect to a popular nonlinear complementary filter in quantitative simulation studies, and we explore how the measurement noise propagates to the asymptotic errors as a function of tuning. \inArxiv{This is an extended version of a paper with the same title (to appear at IFAC WC 2023). Additional mathematical details are provided in this extended version.}\inACC{\vspace*{-20pt}}
\end{abstract}
\if\useieeelayout1
\end{frontmatter}
\fi

\section{Introduction}

The inertial measurement unit (IMU) is a ubiquitous sensor in modern robotics, often used in conjunction with other sensing modalities to infer a system's rotational degrees of freedom. In applications such as micro quadrotor control, it is essential to acquire these estimates at high rates to implement controllers with sufficient bandwidth, necessitating computationally lightweight estimators.

Largely driven by aerospace applications, a significant body of work exists on how to fuse the IMU measurements into an accurate estimate of the rotation and gyroscopic biases, see, e.g.,~\citep{markley2005nonlinear,zamani2015nonlinear,ligorio2015novel,caruso2021analysis}. In the context of attitude estimation, the early work of~\citep{farrell1970attitude} set the grounds for the myriad nonlinear Kalman filters since proposed. These Bayesian methods are often used in practice due to their simplicity and flexibility. However, while the extended, unscented, and other variant assumed Gaussian density filters revert to a standard Kalman filter in a linear setting for which convergence guarantees exist\inArxiv{(see, e.g.,~\citep{sarkka2013bayesian})}, little can be said about worst case performance, convergence, and robustness of these nonlinear filters~\citep{arasaratnam2009cubature}. It is worth noting that Bayesian particle filters
\citep{arulampalam2002tutorial} are asymptotically optimal in the nonlinear setting as the number of particles (and implicitly, the computational burden) approaches infinity. These have also been considered for attitude estimation in~\citep{cheng2004particle}, but are not practical given how fast the estimates need to be computed. Due to the flexibility of these approaches, both \emph{attitude kinematics} and \emph{attitude dynamics} have been considered, often in conjunction with other modalities such as camera and GPS measurements\inArxiv{~\citep{johansen2017nonlinear}}.


An alternative approach is to work with nonlinear stability theory, and not presuppose anything about the noise statistics, but rather design observers which are implicitly robust to disturbances. This method is used in the vast literature on nonlinear complementary filtering, culminating with the seminal works of~\citep{mahony2005complementary,mahony2008nonlinear}. Here, several observers are derived for \emph{attitude kinematics} using Lyapunov theory, with subsequent applications in~\citep{mahony2012multirotor}\inArxiv{ and recent extensions in~\cite{mahony2022observer}}. A similar approach is taken in~\citep{berkane2017design}, where the observer gains are made state dependent to further improve robustness.

However, when considering control applications, we are generally also interested in the attitude rates to compute the actuating torques. An appealing alternative is therefore to consider the \emph{attitude dynamics}, making use of the torques to compute filtered estimates of the attitude, the gyroscopic biases, and the attitude rates. 
Nevertheless, the application of above mentioned methods to \textit{attitude dynamics} is less explored. Some work has been done in, e.g.,~\citep{ng2020secondorder,lu2016gyrofree}, but in these works gyroscopic measurements have been ignored.
To the best knowledge of authors, there exist no works that show uniform local exponential stability and uniform almost global asymptotic stability of the error dynamics in this setting, producing filtered estimates of the attitude, the attitude rate, and the gyroscopic biases. We contribute such a solution, which is important for three reasons: it facilitates the derivation of filtered output feedback controllers for the \emph{attitude dynamics} with explicit gyroscopic bias estimation, permitting extensions of~\citep{lefebergreiffrobertsson2020}. Secondly, the \emph{uniform} stability property provides rigorous robustness guarantees in the sense of~\cite[Lemma~9.3]{Khalil}. Finally, the observer comes with almost global convergence guarantees in contrast to the nonlinear Kalman filters that are often considered for this problem.




\subsection{Outline}
The mathematical preliminaries are given in Sec.~\ref{sec:prelim}, before stating the problem formulation in Sec.~\ref{sec:problem}. The main results are presented in Sec.~\ref{sec:main} in four steps: we (i) start by presenting an observer for the angular momentum in the inertial frame; (ii) restate the seminal result by Mahony; (iii) combine these two observers with a convex combination of the innovation terms; and (iv) describe how the attitude rate estimates can be recovered in the body-fixed frame. This is illustrated by numerical results in Sec.~\ref{sec:numerical}, and the conclusion in Sec.~\ref{sec:final} closes the paper. \inACC{A discrete-time implementation is provided as Matlab code in Appendix~\ref{app:B}, and additional mathematical details and simulation results are given in~\citep{lefeber2023ifac}.}\inArxiv{Some key steps in the proofs are elaborated upon in Appendix~\ref{app:A}, and a discrete-time implementation is provided as Matlab code in Appendix~\ref{app:B}.}

\section{Preliminaries}\label{sec:prelim}
In this section we introduce the notation, definitions and theorems used in the remainder of this paper.

\begin{thm}[{Corollary of \citet[Theorem~1]{loria2005nested}}]
\label{th:matrosov}
Consider the dynamical system
\begin{align}
\label{eq:matsys}
\dot x&=f(t,x) & x(t_0)&=x_0 & f(t,0)&=0,
\end{align}
with $f:\R^+\times\R^n\to\R^n$ locally bounded, continuous and locally uniformly continuous in $t$.

If there exist $j$ differentiable functions $V_i:\R^+\times\R^n\to\R$, bounded in $t$, and continuous functions $Y_i:\R^n\to\R$ for $i\in\{1,2,\dots j\}$ such that
\begin{itemize}
\item $V_1$ is positive definite and radially unbounded,
\item $\dot V_i(t,x)\leq Y_i(x)$, for all $i\in\{1,2,\dots,j\}$,
\item $Y_i(x)=0$ for $i\in\{1,2,\dots,k-1\}$ implies $Y_k(x)\leq 0$, for all $k\in\{1,2,\dots,j\}$,
\item $Y_i(x)=0$ for all $i\in\{1,2,\dots,j\}$ implies $x=0$,
\end{itemize}
then the origin $x=0$ of \eqref{eq:matsys} is uniformly globally asymptotically stable (UGAS).
\end{thm}
For definitions of uniform global (or local) asymptotic (or exponential) stability (UGAS/UGES/ULES), refer to \citep{Khalil}.
\begin{defn}
The origin of \eqref{eq:matsys} is \emph{uniformly almost globally asymptotically stable (UaGAS)} if it is UGAS, except for initial conditions in a set of measure zero.
\end{defn}

We consider rotations $R\in \textup{SO(3)}=\{R\in\R^{3\times 3}\mid R^\top R=I,\det{R}=1\}$, and define the skew-symmetric map
\begin{align}
\label{eq:S}
S(a)=-S(a)^\top =\begin{bmatrix} 0 & -a_3 & a_2\\ a_3 & 0 & -a_1\\ -a_2 & a_1 & 0\end{bmatrix}\in\mathfrak{so}(3).
\end{align}
As the cross product can be expressed $a\times b=S(a)b$, the following useful properties hold for $S:\mathbb{R}^3\mapsto \mathbb{R}^{3\times 3}$:
\inArxiv{\begin{subequations}
\begin{align}
S(a)^\top&=-S(a) & \forall a&\in\R^3\\
S(a)b&=-S(b)a & \forall a,b&\in\R^3\\
a^\top S(b)a&=0 & \forall a,b&\in\R^3\\
RS(a)&=S(Ra)R & \forall R\in\textup{SO(3)}, \forall a&\in\R^3\label{eq:rotid}\\
S(a)S(b)&=ba^\top-(b^\top a)I_3& \forall a,b&\in\R^3.
\end{align}
\end{subequations}}
\inACC{\begin{align*}
S(a)^\top&=-S(a) & \forall a&\in\R^3\\
S(a)b&=-S(b)a & \forall a,b&\in\R^3\\
a^\top S(b)a&=0 & \forall a,b&\in\R^3\\
RS(a)&=S(Ra)R & \forall R\in\textup{SO(3)}, \forall a&\in\R^3\\
S(a)S(b)&=ba^\top-(b^\top a)I_3& \forall a,b&\in\R^3.
\end{align*}}
\noindent\hspace{-7pt} We let $\|x\|_2 = (x^{\top} x)^{1/2}$, using the same notation referring to the induced two norm in the context of matrices. We also consider $\mathcal{L}_2$-norms over an interval $[a,b]$ defined in these norms, as $\|x\|_{\mathcal{L}_2([a,b])}=(\int_{a}^b\|x(t)\|_2^2\mathrm{d} t)^{1/2}$.

\begin{lem}[{\citet[Lemma~5]{lefebergreiffrobertsson2020}}]
\label{lem:ReVdot}
Consider the dynamical systems
$\dot R_1=R_1S(\omega_1)$ and
$\dot R_2=R_2S(\omega_2)$.
Let $R_{12}=R_1R_2^\top$ and $\omega_{12}=\omega_1-\omega_2$. Then
\begin{align}
\label{eq:12dyna}
\dot R_{12}&=R_{12}S(R_2\omega_{12})=S(R_1\omega_{12})R_{12},
\end{align}
and differentiating for some constant vector $v$
\begin{align*}
V=\frac12(R_{12}v-v)^\top (R_{12}v-v)=\frac12\left\|R_{12}v-v\right\|_2^2,
\end{align*}
along solutions of \eqref{eq:12dyna} results in
$\dot V=\omega_{12}^\top S(R_1^\top v)R_2^\top v$.
\end{lem}
\begin{lem}
\label{lem:nul}
Define $r_k=\sum_{i=1}^n k_iS(R^\top v_i)v_i$ with $k_i>0$ and $v_i\in\mathbb{R}^3$ such that $M=\sum_{i=1}^nk_iv_iv_i^\top=U\Lambda U^\top$ with $U\in\textup{SO(3)}$ and
$\Lambda$ a diagonal matrix with distinct eigenvalues $\lambda_i$, i.e., $\lambda_3>\lambda_2>\lambda_1>0$.
Then $r_k=0$ implies that $U^\top RU\in\{I,D_1,D_2,D_3\}$, where $D_1=\diag(1,-1,-1)$, $D_2=\diag(-1,1,-1)$, $D_3=\diag(-1,-1,1)$.
Furthermore, if in addition $\dot R=RS(\omega)$ and $\dot  r_k=0$, then also $\omega=0$.
\end{lem}
\begin{pf}
The first claim was shown in \citep{mahony2008nonlinear}. By defining $\bar R=URU^\top$, $\bar\omega=U\omega$, and $\bar k_i=k_iv_i^\top v_i$ it follows that \inACC{w.l.o.g.} \inArxiv{without loss of generality,} we can assume that $U=I$ and $v_i^\top v_i=1$.
Then $r_k=0$ implies $R=\diag(r_1,r_2,r_3)=\diag(\pm 1,\pm 1, \pm 1)$. Let $\Lambda=\diag(\lambda_1,\lambda_2,\lambda_3)$. Then we have
\begin{align}
\dot  r_k
&= -\!\sum_{i=1}^nk_iS(S(\omega) R^\top v_i) v_i\notag\arxivbreak=-\!\sum_{i=1}^nk_iS( v_i)S( R^\top v_i)\omega\notag\\
&=-
\diag(r_2\lambda_2+r_3\lambda_3,r_1\lambda_1+r_3\lambda_3,r_1\lambda_1+r_2\lambda_2)
\omega,\label{eq:rkdot}
\end{align}
from which we can conclude that $\dot  r_k=0$ implies $\omega=0$, since $\lambda_i$ are distinct and $r_i\in\{-1,1\}$.
\end{pf}

\section{Problem formulation}\label{sec:problem}
Let $R\in \textup{SO(3)}$ denote the rotation matrix from the body-fixed frame to the inertial frame and let $\omega\in\R^3$ denote the body-fixed angular velocities.
Then the \emph{kinematics} of a rotating rigid body can be described by
\begin{align}
\label{eq:kinematics}
\dot R &= RS(\omega),
\end{align}
where $\omega$ is regarded as input.
Consider the outputs
\begin{align}
\label{eq:output}
y_0&=\omega+b & y_i&=R^\top v_i & i=1,\dots, n,
\end{align}
where $b$ is an unknown constant, and $v_i$ denote $n$ known inertial directions.
That is, assume biased measurement of angular velocities and body-fixed frame observations of the fixed inertial directions $v_i$.
\begin{assumption}
For attitude reconstruction $n\geq 2$ independent inertial directions are required. However, if we have two independent directions $v_1$ and $v_2$, then $v_3=v_1\times v_2=S(v_1)v_2$ is a third independent direction. Therefore, in the remainder we assume \inACC{w.l.o.g.} \inArxiv{without loss of generality} that $n\geq 3$ instead.
\end{assumption}
In this setting, a large number of observers exist, such as the filters in the seminal work of ~\citep{mahony2008nonlinear}:
\begin{thm}[{\citet[Th.~5.1]{mahony2008nonlinear}}]
Consider
the explicit complementary filter with bias correction
\begin{align}
\label{eq:Mahony}
\dot{\hat b}&=k_b\tilde r_k &
\dot{\hat R}&=\hat RS(y_0-\hat b-k_R\tilde r_k), &
\end{align}
where $\tilde r_k=\sum_{i=1}^n k_iS(\hat R^\top v_i)y_i$, $k_R>0$, and $k_b>0$. Define the estimation errors
$\tilde R=\hat RR^\top$ and $\tilde b=\hat b-b$.
If $\omega(t)$ is a bounded absolutely continuous signal, the pair of signals $(\omega(t),\tilde R)$
is asymptotically independent, and the weights $k_i>0$ are chosen such that
$M=\sum_{i=1}^nk_iv_{i}{v_i}^\top$
has distinct eigenvalues, then $(\tilde R,\tilde b)$ is almost globally asymptotically stable
and locally exponentially stable to $(I,0)$.
\end{thm}
This explicit complementary filter with bias correction \eqref{eq:Mahony} has seen much use in practice.
However, this filter only produces estimates for the attitude and bias, but not an estimate for the angular velocities.
Clearly, from measurements $y_0$ and bias estimate $\hat b$ an unbiased estimate for the angular velocities is available, but for noisy $y_0$ this unbiased estimate for the angular velocities is also noisy and not a filtered signal.
Therefore, the goal of this paper is to extend the explicit complementary filter with bias correction to the \emph{dynamics} of a rotating body, producing not only filtered estimates for the attitude and bias, but also filtered unbiased estimates for the angular velocities. To be precise, we aim to solve the following problem.
\begin{problem}
\label{problem}
The motion of a rotating rigid body configured on $R\in\textup{SO(3)}$ is governed by the dynamics
\label{eq:dynamics}
\begin{align}
\dot R &= RS(\omega)&
J\dot\omega &= S(J\omega)\omega+\tau,
\end{align}
where $J=J^\top >0$ denotes the inertia matrix with respect to the body-fixed frame and $\tau\in\R^3$ denotes the total moment vector in the body-fixed frame, is a known input.

Consider the outputs \eqref{eq:output}. Design an observer/filter which produces estimates $\hat R$, $\hat\omega$, and $\hat b$ such that the point $(I,0,0)$ of the estimation error dynamics $(\tilde R,\tilde\omega,\tilde b)$,
given by
\begin{align}
\label{eq:error}
\tilde R&=\hat RR^\top & \tilde\omega&=\hat\omega-\omega & \tilde b&=\hat b-b,
\end{align}
is almost globally and locally exponentially stable.
\end{problem}

\section{Main results}\label{sec:main}
The difficulty in almost globally solving Problem~\ref{problem} is dealing with the Coriolis-terms, which contains quadratic expressions in the angular velocities.
Our way around this difficulty is to first design an observer for the angular \emph{momentum} expressed in the \emph{inertial frame}. Next, our estimate for the attitude can be used to transform those estimates into estimates for the angular \emph{velocities} expressed in the \emph{body-fixed frame}.

As a first step, we consider the problem of designing an observer for both the attitude and the angular momentum expressed in the inertial frame without using the measurement of angular velocities. As a second step, we revisit the explicit complementary filter with bias correction by \citep{mahony2008nonlinear} to prepare for our third step.
In our third step we fuse the observers derived in the previous steps to produce an estimates for the attitude, the angular momentum expressed in the inertial frame, and a bias estimate. In our fourth and final step, the derived estimates are used to estimate the angular velocities in the body-fixed frame using only the measured outputs in~\eqref{eq:output}.

\subsection{Step 1: Angular momentum estimator}
Our first goal is to design an observer for estimating the angular momentum expressed in the inertial frame using only the body-fixed frame observations of fixed inertial directions, that is without using measurement of angular velocities.
To that end, define $\ell =RJ\omega$, so $\omega=J^{-1}R^\top \ell$.
Then we get as resulting dynamics:
\begin{align}
\label{eq:dynamics1}
\dot R&=RS(J^{-1}R^\top \ell) & \dot\ell&=R\tau.
\end{align}
Consider only the outputs
\begin{align}
y_i&=R^\top v_i & i=1,\dots, n.
\end{align}
Our goal is to construct estimates $\hat R$ and $\hat\ell$ such that the estimation errors
\begin{align}
\label{eq:error1}
\tilde R&=\hat RR^\top &
\tilde\ell&=\hat\ell-\ell
\end{align}
converge to $I$ respectively $0$. Define the following observer:
\begin{subequations}
\label{eq:observer1}
\begin{align}
\dot{\hat R}&=\hat RS\left(J^{-1}R^\top \hat\ell-k_R\tilde r_k\right)&
\dot{\hat\ell}&=R\tau-k_\ell RJ^{-1}\tilde r_k,
\end{align}
where $k_R>0$, $k_\ell>0$, and
\begin{align}
\label{eq:tilderk}
\tilde r_k&=\sum_{i=1}^n k_iS(\hat R^\top v_i)R^\top v_i=\sum_{i=1}^n k_iS(\hat R^\top v_i)y_i.
\end{align}
\end{subequations}
\begin{prop}
\label{prop:1}
Consider the observer \eqref{eq:observer1} in closed-loop with the dynamics \eqref{eq:dynamics1}. If $\omega$ and $\dot\omega$ are bounded and the weights $k_i$ are chosen such that $\sum_{i=1}^nk_iv_iv_i^\top $ has distinct eigenvalues $\lambda_i$, i.e., $\lambda_3>\lambda_2>\lambda_1>0$, then the estimation errors \eqref{eq:error1} are UaGAS and ULES towards $(I,0)$.
\end{prop}
\begin{pf}\label{pf:2}
Using Lemma~\ref{lem:ReVdot}, the estimation error dynamics can be written as
\begin{subequations}
\label{eq:errordyna1}
\begin{align}
\dot{\tilde R}&=\tilde RS\Bigl(R\bigl[J^{-1}R^\top \tilde\ell-k_R\tilde r_k\bigr]\Bigr)\label{eq:Rtildedot1}\\
\dot{\tilde\ell}&=-k_\ell RJ^{-1}\tilde r_k.
\end{align}
\end{subequations}
Differentiating the Lyapunov function candidate
\begin{align}
V_1&=k_\ell\sum_{i=1}^n\myfrac{k_i}{2}\|\tilde Rv_i-v_i\|_2^2+\myfrac{1}{2}\tilde\ell^\top \tilde\ell,
\end{align}
along \eqref{eq:errordyna1}, using Lemma~\ref{lem:ReVdot}, results in
\begin{align}
\dot V_1&=k_\ell(RJ^{-1}R^\top \tilde\ell-k_R\tilde r_k)^\top \tilde r_k
+\tilde\ell^\top [-k_\ell RJ^{-1}R^\top \tilde r_k]\notag\\
&=-k_\ell k_R\left\|\tilde r_k\right\|_2^2=Y_1,
\end{align}
which is negative semi-definite. Differentiating
$V_2=-\tilde r_k^\top\dot{\tilde r}_k$
along \eqref{eq:errordyna1} results in
\begin{subequations}
\begin{align}
\dot V_2&=-\|\dot{\tilde r}_k\|_2^2-\tilde r_k^\top\ddot{\tilde r}_k\arxivbreak \leq -\|\dot{\tilde r}_k\|_2^2+K\|\tilde r_k\|_2\label{eq:p2firstbound}\\
&\leq-\gamma\|\tilde\ell\|_2^2+\bar K\|\tilde r_k\|_2=Y_2.\label{eq:p2secondbound}
\end{align}
\end{subequations}
The first inequality follows from boundedness of $\ddot{\tilde r}_k$ which follows from $\dot V_1\leq 0$ and \eqref{eq:errordyna1}. The second inequality follows from \eqref{eq:rkdot} and \eqref{eq:Rtildedot1}.
Applying Theorem~\ref{th:matrosov} shows UGAS towards $\tilde r_k=0$, $\dot r_k=0$, which, using Lemma~\ref{lem:nul} implies UaGAS towards $\tilde R=I$, $\tilde\ell=0$.
Considering $V_1+\epsilon V_2$, ULES can be shown along the lines of \citep{wulee2016}.
\hfill $\qed$
\end{pf}
\subsection{Step 2: Gyroscopic bias estimator}
As a second ingredient we need the observer of \eqref{eq:Mahony}.
Consider the kinematics \eqref{eq:kinematics} with outputs \eqref{eq:output}. Our goal is to obtain estimates $\hat R$ and $\hat b$ such that the errors
\begin{align}
\label{eq:error2}
\tilde b&=\hat b-b&
\tilde R&=\hat RR^\top
\end{align}
converge to $0$ and $I$, respectively.

Define the following observer/filter:
\begin{align}
\label{eq:observer2}
\dot{\hat b}&=k_b\tilde r_k &
\dot{\hat R}&=\hat RS(y_0-\hat b-k_R\tilde r_k)
\end{align}
with $k_b>0$, $k_R>0$, $J=J^\top >0$ and $\tilde r_k$ as in \eqref{eq:tilderk}.
\begin{prop}
\label{prop:2}
Consider the observer \eqref{eq:observer2} in closed-loop with the kinematics \eqref{eq:kinematics}. If $\omega$ and $\dot\omega$ are bounded and the weights $k_i$ are chosen such that $\sum_{i=1}^nk_iv_iv_i^\top $ has distinct eigenvalues $\lambda_i$, i.e., $\lambda_3>\lambda_2>\lambda_1>0$, then the estimation errors \eqref{eq:error2} are UaGAS and ULES towards $(I,0)$.
\end{prop}
\begin{pf}
The estimation error dynamics are given by
\begin{align}
\label{eq:errordyna2}
\dot{\tilde b}&=k_b\tilde r_k &
\dot{\tilde R}&=\tilde RS(R[-\tilde b-k_R\tilde r_k]).
\end{align}
Differentiating the Lyapunov function candidate
\begin{align}
V_1&= k_b\sum_{i=1}^n\tfrac{k_i}{2}\left\|\tilde Rv_i-v_i\right\|_2^2+\tfrac{1}{2}\tilde b^\top\tilde b
\end{align}
along \eqref{eq:errordyna2} results in
\begin{align}
\dot V_1&= k_b\left(-\tilde b-k_R\tilde r_k\right)^\top\!\!\tilde r_k+\tilde b^\top k_b\tilde r_k=-k_bk_R\left\|\tilde r_k\right\|_2^2,
\end{align}
which is negative semi-definite. Differentiating
$V_2=-\tilde r_k^\top\dot{\tilde r}_k$
along \eqref{eq:errordyna1} results in
\begin{subequations}
\begin{align}
\dot V_2&=-\|\dot{\tilde r}_k\|_2^2-\tilde r_k^\top\ddot{\tilde r}_k\arxivbreak\leq -\|\dot{\tilde r}_k\|_2^2+K\|\tilde r_k\|_2\notag\\
&\leq-\gamma\|\tilde b\|_2^2+\bar K\|\tilde r_k\|_2=Y_2.
\end{align}
\end{subequations}
The first inequality follows from boundedness of $\ddot{\tilde r}_k$ which follows from $\dot V_1\leq 0$, \eqref{eq:rkdot}, and boundedness of $\omega$ and $\dot\omega$. The second inequality follows from \eqref{eq:rkdot} and \eqref{eq:errordyna2}.
The proof can be completed along the lines of that of Proposition~\ref{prop:1}.
\qed
\end{pf}

\begin{rem}
Note that in our proof we do not require that the pair of signals $(\omega(t),\tilde R)$ is asymptotically independent, which is difficult to check since $\tilde R$ is not an external signal (as it is generated in closed-loop with the observer). On the other hand, we need to assume that $\dot\omega$ is bounded, which is a slightly stronger condition than assuming that $\omega$ is absolutely continuous. However, this allows us
to conclude uniform stability, which implies robustness against bounded disturbances by~\cite[Lemma~9.3]{Khalil}.
\end{rem}

\subsection{Step 3: Fusing the two observers}
Our next step is to fuse the two observers \eqref{eq:observer1} and \eqref{eq:observer2} into one.
The observer \eqref{eq:observer1} provides us with an estimate $\hat l$ for the angular momentum expressed in the inertial frame. Therefore, we can consider $J^{-1}R^\top \hat\ell$ as an estimate for the angular velocity.
The observer \eqref{eq:observer2} provides us with a bias estimate so that $y_0-\hat b$ can also be considered as an estimate for the angular velocity. In our combined observer we fuse those to estimates, by using a fraction $\alpha$ of the first estimator, and a fraction $1-\alpha$ of the second estimator.

With this intuition, consider the dynamics \eqref{eq:dynamics1} together with the outputs \eqref{eq:output}. We propose the following observer
\begin{subequations}
\label{eq:observerfused}
\begin{align}
\dot{\hat b}&=k_b\tilde r_k-\alpha k_bk_\alpha J\tilde\delta_L\\
\dot{\hat R}&=\hat RS\left(\alpha J^{-1}R^\top\hat\ell-(1-\alpha )(y_0-\hat b)-k_R\tilde r_k\right)\\
\dot{\hat\ell}&=R\tau-k_\ell RJ^{-1}\tilde r_k-(1-\alpha )k_\ell k_\alpha R\tilde\delta_L,
\intertext{where}
\tilde\delta_L&=R^\top \hat\ell-J(y_0-\hat b)=R^\top\tilde\ell+J\tilde b\label{eq:tildeL}.
\end{align}
\end{subequations}
with $k_\alpha>0$, $k_b>0$, $k_R>0$, $k_\ell>0$, $0<\alpha<1$, and $\tilde r_k$ as defined in \eqref{eq:tilderk}.
\begin{rem}
Note that, $\tilde\delta_L$ can be interpreted as the difference between two estimators for the angular momentum expressed in the body-fixed frame.
\end{rem}

\begin{prop}
\label{prop:3}
Consider the observer \eqref{eq:observerfused} in closed-loop with the dynamics \eqref{eq:dynamics1}. If the weights $k_i$ are chosen such that $\sum_{i=1}^nk_iv_iv_i^\top $ has distinct eigenvalues $\lambda_i$, i.e., $\lambda_3>\lambda_2>\lambda_1>0$, then the estimation errors
\begin{align}
\label{eq:error3}
\tilde R&=\hat RR^\top & \tilde\ell&=\hat\ell-\ell & \tilde b&=\hat b-b,
\end{align}
are UaGAS and ULES towards $(I,0,0)$.
\end{prop}
\begin{pf}
The estimation error dynamics are given by
\begin{subequations}
\label{eq:errordyna3}
\begin{align}
\dot{\tilde b}&=k_b\tilde r_k-\alpha k_bk_\alpha J\tilde\delta_L \\
\dot{\tilde R}&=\tilde RS\Bigl(R\bigl[\alpha J^{-1}R^\top\tilde\ell-(1-\alpha )\tilde b-k_R\tilde r_k\bigr]\Bigr)\\
\dot{\tilde\ell}&=-k_\ell RJ^{-1}\tilde r_k-(1-\alpha )k_\ell k_\alpha R\tilde\delta_L.
\end{align}
\end{subequations}

Differentiating the Lyapunov function candidate
\begin{align}\label{eq:prop3lyap}
V_1&=k_\ell k_b\sum_{i=1}^N\myfrac{k_i}{2}\|\tilde Rv_i-v_i\|_2^2+\myfrac{k_\ell}{2}(1-\alpha )\tilde b^\top \tilde b+\myfrac{k_b}{2}\alpha \tilde\ell^\top \tilde\ell,
\end{align}
along \eqref{eq:errordyna3} results in
\begin{align}
\dot V_1&=k_\ell k_b\left(\alpha J^{-1}R^\top\tilde\ell-(1-\alpha )\tilde b-k_R\tilde r_k\right)^\top \tilde r_k\if\useieeelayout1\notag\fi\\
&\quad+(1-\alpha )k_\ell\tilde b^\top[k_b\tilde r_k-\alpha k_bk_\alpha J\tilde\delta_L ]\if\useieeelayout1\notag\fi\\
&\quad+\alpha k_b\tilde\ell^\top[-k_\ell RJ^{-1}\tilde r_k-(1-\alpha )k_\ell k_\alpha R\tilde\delta_L]\if\useieeelayout1\notag\fi\\
&=-k_\ell k_bk_R\left\|\tilde r_k\right\|_2^2-\alpha (1-\alpha )k_\ell k_bk_\alpha \|\tilde\delta_L\|_2^2,\label{eq:prop3lyapderiv}
\end{align}
which is negative semi-definite. Here we used \eqref{eq:tildeL}.
Differentiating
$V_2=-\tilde r_k^\top\dot{\tilde r}_k$
along \eqref{eq:errordyna3} results in
\begin{align}
\dot V_2&=-\|\dot{\tilde r}_k\|_2^2-\tilde r_k^\top\ddot{\tilde r}_k\notag\arxivbreak\leq -\|\dot{\tilde r}_k\|_2^2+K\|\tilde r_k\|_2\notag\\
&\leq-\gamma\|-R\tilde b+\alpha RJ^{-1}\tilde\delta_L \|_2+\bar K\|\tilde r_k\|_2^2\notag\\
&\leq-\gamma\|\tilde b\|_2+\bar{\bar K}(\|\tilde\delta_L\|_2+\|\tilde r_k\|_2)=Y_2
\end{align}
The proof can be completed along the lines of that of Proposition~\ref{prop:1}. \hfill $\qed$
\end{pf}
\begin{rem}
Note that, like in Proposition~\ref{prop:1}, there is no need for assuming that $\omega$ or $\dot\omega$ (or $\tau$) are bounded. From $\dot V_1\leq 0$ we have boundedness of the estimation errors, which is all we need to complete the proof.
\end{rem}
\begin{rem}
Note that for $\alpha=0$ or $\alpha=1$ the observer \eqref{eq:observerfused} reduces to respectively \eqref{eq:observer1} or \eqref{eq:observer2}, for which we obtained results in Proposition~\ref{prop:1} respectively Proposition~\ref{prop:2}.
\end{rem}

\subsection{Step 4: Final result}
Our final step is to replace the estimate $\hat\ell$ for the angular momentum expressed in the inertial frame, obtained from the observer \eqref{eq:observerfused}, by a filtered estimate $\hat\omega$
for the angular velocity expressed in the body-fixed frame. Furthermore, we need to overcome the problem that we do not know $R$, which is used in \eqref{eq:observerfused}, as we only have
\eqref{eq:output} available for measurement, not $R$ itself.

The latter is actually less of a problem than it might seem at first glance.
We assumed that the weights $k_i$ are chosen such that $M=\sum_{i=1}^nk_iv_iv_i^\top=U\Lambda U^\top$ has distinct eigenvalues  $\lambda_i$, i.e., $\lambda_3>\lambda_2>\lambda_1>0$. Therefore, the matrix $M$ is invertible and we obtain
\begin{align}
\label{eq:Rrewrite}
R=M^{-1}\sum_{i=1}^nk_iv_iv_i^\top R=M^{-1}\sum_{i=1}^nk_iv_iy_i^\top.
\end{align}
As a result, each occurrence of $R$ in \eqref{eq:observerfused} can be replaced by the right hand side of \eqref{eq:Rrewrite}. We emphasize that~\eqref{eq:Rrewrite} \emph{is not} the attitude estimate, the attitude estimate $\hat{R}$ is still computed and updated through the ODEs in~\eqref{eq:observerfused}.

Our filtered estimate for the angular velocity expressed in the body-fixed frame is given by
$\hat\omega=J^{-1}\hat R^\top \hat\ell$.
We can now summarize our result in the following.
\begin{prop}\label{prop:4}
Consider the dynamics \eqref{eq:dynamics} and output \eqref{eq:output} in closed-loop with the observer
\begin{subequations}
\label{eq:observerfinal}
\begin{align}
\dot{\hat b}&=k_b\tilde r_k-\alpha k_bk_\alpha J\tilde\delta_L\\
\dot{\hat R}&=\hat RS\left(\alpha J^{-1}\tilde\delta_L+y_0-\hat b-k_R\tilde r_k\right)\label{eq:dotRprop4}\\
\dot{\hat\ell}&=\bar R[\tau-k_\ell J^{-1}\tilde r_k-(1-\alpha )k_\ell k_\alpha\tilde\delta_L]\\
\hat\omega&=J^{-1}\hat R^\top \hat\ell
\intertext{where}
\tilde r_k&=\sum_{i=1}^n k_iS(\hat R^\top v_i)y_i\\
\tilde\delta_L&=\bar R^\top \hat\ell-J(y_0-\hat b)\\
\bar R&=\left(\sum_{i=1}^nk_iv_iv_i^\top\right)^{-1}\sum_{i=1}^nk_iv_iy_i^\top.
\end{align}
\end{subequations}
Let $k_\alpha>0$, $k_b>0$, $k_R>0$, $k_\ell>0$, $0<\alpha<1$.
If in addition $k_i$ are chosen such that $M=\sum_{i=1}^nk_iv_iv_i^\top=U\Lambda U^\top$ has distinct eigenvalues $\lambda_i$, i.e., $\lambda_3>\lambda_2>\lambda_1>0$, the
the observer errors \eqref{eq:error} are UaGAS and ULES towards $(I,0,0)$, provided that $\omega$ is bounded.
\end{prop}
\begin{pf}
From Proposition~\ref{prop:3} we have that $\tilde R$, $\tilde b$ and $\tilde\ell$ are UaGAS and ULES towards $(I,0,0)$. Therefore, it only remains to show convergence of $\tilde\omega$. We have
\begin{align}
\tilde\omega&=J^{-1}\hat R^\top\hat\ell-\omega=\underbrace{J^{-1}\hat R^\top\tilde\ell}_{\to 0}-\underbrace{J^{-1}R^\top[\tilde R-I]RJ}_{\to 0}\omega,
\end{align}
which explains the additional requirement that $\omega$ is bounded, in comparison with Proposition~\ref{prop:3}.
\hfill $\qed$
\end{pf}

\section{Numerical examples}\label{sec:numerical}
In this section, we present three numerical examples. The first is a qualitative simulation to illustrate typical convergence behaviors of the estimators. Next, we give quantitative results showing the utility of combining the observers as in Propositions~\ref{prop:3}--\ref{prop:4} by studying the statistics of the transient and stationary errors. Finally, we discuss how to tune the observers based on the asymptotic errors, and how these errors are affected by measurement noise.

\subsection{Typical convergence in an ideal setting}\label{sec:ideal}
In this ideal setting, we take the measurements to be noise-free and initialize a simulation with initial errors and parameters sampled from the distributions in Appendix~\ref{sec:parameters}. The dynamical system in~\eqref{eq:dynamics} is driven by a torque sequence
\begin{equation}
    \tau(t) = (\sin(t+1), \sin(2t+2), \sin(3t+3))^{\top}\in\mathbb{R}^3,
\end{equation}
where the initial conditions and parameters are realized as
\begin{align*}
R(0) &
\hspace{-2.5pt}=\hspace{-3.5pt}
\begin{bmatrix}
    0.18 &    0.97 &   -0.15\\
    0.08 &    0.14 &    0.99\\
    0.98 &   -0.19 &   -0.06
\end{bmatrix}\hspace{-3pt},&\hspace{-7pt}\omega(0)&
\hspace{-2.5pt}=\hspace{-3.5pt}
\begin{bmatrix}
   -0.11\\
    0.02\\
   -0.06
\end{bmatrix}\hspace{-3pt},&\hspace{-7pt}
{b} &
\hspace{-2.5pt}=\hspace{-3.5pt}
\begin{bmatrix}
   -0.12\\
   -2.54\\
    0.28
\end{bmatrix}\hspace{-3pt},\\
\hat{R}(0) &
\hspace{-2.5pt}=\hspace{-3.5pt}
\begin{bmatrix}
    0.35 &    0.06 &    0.94\\
    0.84 &    0.42 &   -0.34\\
   -0.41 &    0.91 &    0.09
\end{bmatrix}\hspace{-3pt},&
\hspace{-7pt}\hat{\ell}(0) &
\hspace{-2.5pt}=\hspace{-3.5pt}
\begin{bmatrix}
   -1.12\\
    0.05\\
   -1.24
\end{bmatrix}\hspace{-3pt},&
\hspace{-7pt}\hat{b}(0) &
\hspace{-2.5pt}=\hspace{-3.5pt}
\begin{bmatrix}
   -0.83\\
    0.54\\
    0.11
\end{bmatrix}\hspace{-3pt},\\
J &
\hspace{-2.5pt}=\hspace{-3.5pt}
\begin{bmatrix}
    0.91 &   0.03 &   0.14\\
    0.03 &   0.73 &   0.15\\
    0.14 &   0.15 &   0.64\\
\end{bmatrix}\hspace{-3pt},&&&&\hspace{-98pt}\inArxiv{\hspace{-25pt}}
\begin{bmatrix}
v_1 & \hspace{-2pt}v_2 & \hspace{-2pt}v_3
\end{bmatrix}
\hspace{-2.5pt}=\hspace{-3.5pt}
\begin{bmatrix}
         0.00 &  -0.87&   -0.45\\
         0.00 &  -0.50&    0.87\\
        -1.00 &  -0.05&    0.00
\end{bmatrix}\hspace{-3pt}\inACC{.}\inArxiv{,}
\end{align*}
\inArxiv{here rounded to two decimals to ease visualization}. In this example, we tune the observer with
\begin{subequations}
\begin{align}
    k_R&=2.0,&
    k_l&=2.0,&
    k_a&=1.0,&
    k_b&=4.0,\\
    k_1&=1.1,&
    k_2&=1.2,&
    k_3&=1.3,&
    \alpha&=0.3.
\end{align}
\end{subequations}
\begin{figure}[t!]
    \centering
    \inACC{\includegraphics[width=\columnwidth]{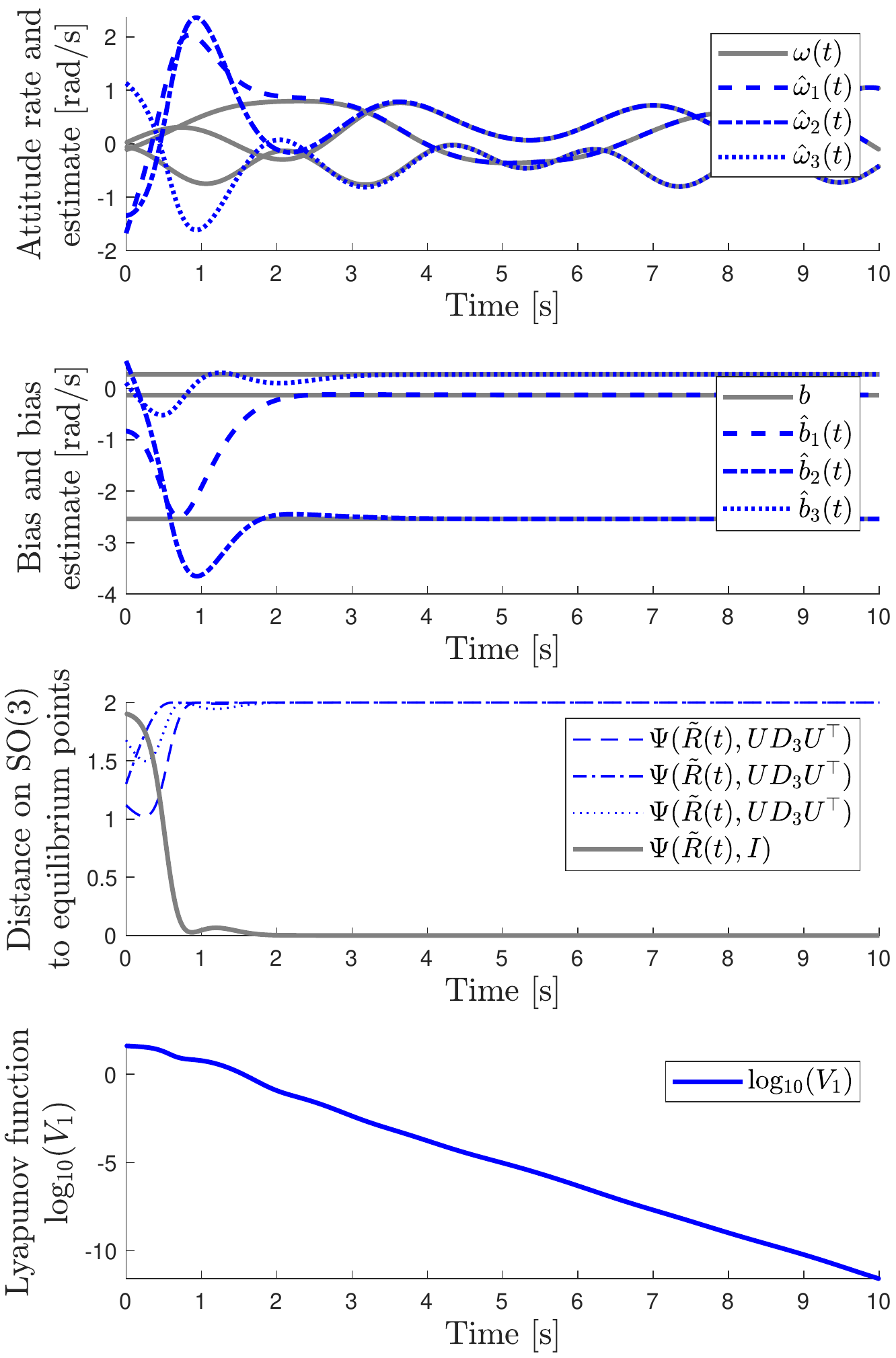}\vspace*{-15pt}}
    \inArxiv{\includegraphics[width=0.6\textwidth]{figures/example1new.pdf}}
    \caption{State trajectory (gray) and estimated states (blue), along with error signals. The gyroscopic biases, and attitude rates are shown in the top two subplots, respectively. The distance of the estimation error $\tilde{R}$ to the four equilibrium points is shown in the third subplot, with $\Psi(\tilde{R},I)$ in gray. The Lyapunov function is depicted in the 10-logarithm in the bottom subplot.}
    \label{fig:example1qualitative}
\end{figure}%
This results in a matrix $M$ in~\eqref{eq:Rrewrite} with distinct eigenvalues $\Lambda=\diag(1.07, 1.23, 1.30)$. The effects of the observer tuning are discussed later in Sec.~\ref{sec:numtuning}. The resulting system response is shown in Fig.~\ref{fig:example1qualitative}, where $\Psi(A,B)=\tfrac{1}{2}\mathrm{Tr}(A^{\top}B - I)$. Despite initializing the estimator very away from the stable equilibrium point in this measure, we obtain a good estimate within seconds with relatively small transients in the attitude rate and bias estimates. For small errors, we observe a linear decay of the Lyapunov function $V_1$ in~\eqref{eq:prop3lyap} in the 10-logarithm, as expected from the ULES property. 

\subsection{Quantitative Monte Carlo results with noise}
One of the more important effects of having $\alpha\in(0,1)$ is that we effectively filter both the bias and the attitude rates, which reduces the impact of the noise in these estimates. To quantify and demonstrate this, we consider the same tuning as in Sec.~\ref{sec:ideal}, and compute the root mean-square error (RMSE) of the $\mathcal{L}_2$-norms in the signals $\|\tilde\omega(t)\|_2$, $\|\tilde{b}(t)\|_2$, and $\Psi(\tilde{R}(t),I)$. That is, we consider $N_{MC}$ realizations of the parameters in Appendix~\ref{sec:parameters}, denote a trajectory from the $i\mathrm{th}$ simulation as $x^{(i)}(t)$, and let
\begin{align}
\mathrm{RMSE}_{\mathcal{L}_2([{a,b}])}(x)&=\Big(\frac{1}{{N_{MC}}}\sum_{i=1}^{N_{MC}} {\int_{a}^{b}}\|x^{(i)}(t)\|_2^2\mathrm{d} t\Big)^{1/2}\hspace{-4pt}.
\end{align}
Here, by considering this measure over the entire simulation time, $t\in[0,T]$, we capture the length of the initial transients, and by considering it over the last second of the simulation, $t\in[T-1,T]$, we capture the stationary errors primarily induced by the noise. These measures are shown in Table~\ref{tab:tabexample3}, as computed from ${N_{MC}}=10^{3}$ realizations.
\begin{table}[h!]
    \centering
    \caption{RMSEs of transient and stationary errors categorized by signals and observers.}
    \begin{tabular}{c@{\hspace{4pt}}|@{\hspace{4pt}}c@{\hspace{4pt}}|@{\hspace{4pt}}c@{\hspace{4pt}}|@{\hspace{4pt}}c@{\hspace{4pt}}|@{\hspace{4pt}}c@{\hspace{4pt}}|@{\hspace{4pt}}c@{\hspace{4pt}}|@{\hspace{4pt}}c}\toprule
        Measure &\multicolumn{3}{c}{$\mathrm{RMSE}_{\mathcal{L}_2([0,T])}(x)$}&\multicolumn{3}{@{\hspace{-4.3pt}}|c}{$\mathrm{RMSE}_{\mathcal{L}_2([T-1,T])}(x)$}\\\midrule
        Signal & $\Psi(\tilde{R},I)$ & $\tilde{\omega}$& $\tilde{b}$
        & $\Psi(\tilde{R},I)$ & $\tilde{\omega}$& $\tilde{b}$\\
        \midrule\midrule
            Prop.~\ref{prop:1} & 0.560&    2.571&    2.629&    3.043$\cdot 10^{-5}$&    0.022&    0.178\\
            Prop.~\ref{prop:2} & 0.577&    2.463&    2.401&    2.718$\cdot 10^{-5}$&    0.177&    0.016\\
            Prop.~\ref{prop:4} & 0.570&    2.389&    2.226&    2.809$\cdot 10^{-5}$&    0.021&    0.016\\
        \bottomrule
    \end{tabular}
    \label{tab:tabexample3}
\end{table}

\begin{rem}
Here we note that there is significant variance in these measures when considered over the entire simulation time (i.e., with $[0,T]$), but the standard deviation of $\mathrm{RMSE}_{\mathcal{L}_2([T-1,T])}(x)$ is in the order of $10^{-10}$ for $\Psi(\tilde{R},I)$, and the order of $10^{-4}$ for $\tilde{\omega}$ and $\tilde{b}$, respectively. As such, there is a statistically significant difference in stationary performance between the observers when considering the parameter, noise, and error distributions in Appendix~\ref{sec:parameters}.
\end{rem}

From these results, we note that the transient responses are similar in the three observers, but that the stationary noise levels differ greatly. In particular, the observer in Proposition~\ref{prop:1} achieves low noise levels in the attitude rate errors, as the attitude rate estimate is filtered in the observer, but the stationary noise in the bias is relatively large. For the observer in Proposition~\ref{prop:2}, the relationship is the reverse. Finally, for the observer in Proposition~\ref{prop:4}, we filter both signals, resulting in low noise levels both in the attitude rate error and in the bias. In this simulation study, the asymptotic noise levels differ by almost one magnitude. If the observer is to be used for feedback control on the estimates $(\hat{R},\hat{\omega})$ based on noisy measurements $\{y_i\}_{i=0}^n$, it is clear that the observers in Proposition~\ref{prop:1} and Proposition~\ref{prop:4} should be considered over Proposition~\ref{prop:2} (the result of~\citep{mahony2008nonlinear}). Additionally, we note that there is clear merit to considering Proposition~\ref{prop:4} over Proposition~\ref{prop:1} if the asymptotic noise in the bias estimates are of concern.

\subsection{Observer tuning}\label{sec:numtuning}
The tuning of the estimator is non-trivial, and somewhat counter intuitive. Some insight can be gained by following~\protect{\cite[Section 5.4]{greiff2021nonlinear}} and taking a local approximation of the attitude error close to the identity element, $\tilde{R} = I + S(\tilde{\epsilon}) + o(\|\tilde{\epsilon}\|_2^2)$. Here, we define measurement noise as an additive perturbation on $y_0$, and a multiplicative disturbance on $\{y_i\}_{i=1}^n$ perturbing the direction, with
\begin{equation}
    y_0 = w + b + \delta_0, \qquad
    y_i =R^{\top}(I + S(\delta_i))v_i.
\end{equation}
We then express the local error dynamics in~\eqref{eq:errordyna3} in $\tilde{x}^{\top} = (\tilde{\epsilon}^{\top}, \tilde{\omega}^{\top}, \tilde{b}^{\top})\in\mathbb{R}^9$, driven by $\delta^{\top} = (\delta_0^{\top}, \delta_1^{\top}, \delta_2^{\top}, \delta_3^{\top})\in\mathbb{R}^{12}$, and linearize the system about the origin, resulting in
\begin{equation}\label{eq:linsysest}
\dot{\tilde{x}} = A\tilde{x} + B\delta.
\end{equation}
Here, we compute $(A,B)$ using the automatic differentiation tool \texttt{CasADi} in~\citep{andersson2012casadi}. This permits us to study how the tuning of the estimator affects the properties of the linear system in~\eqref{eq:linsysest} governing the local estimation errors, and also facilitates reasoning about how certain noises affect the stationary errors by \inArxiv{tools from linear systems theory, such as the singular-value plots}\inACC{computing the singular-values} from the inputs $\delta_i$ to the errors $\tilde{x}$.

\newcommand{\sigmaplot}{
    \centering
    \inACC{\includegraphics[width=\columnwidth]{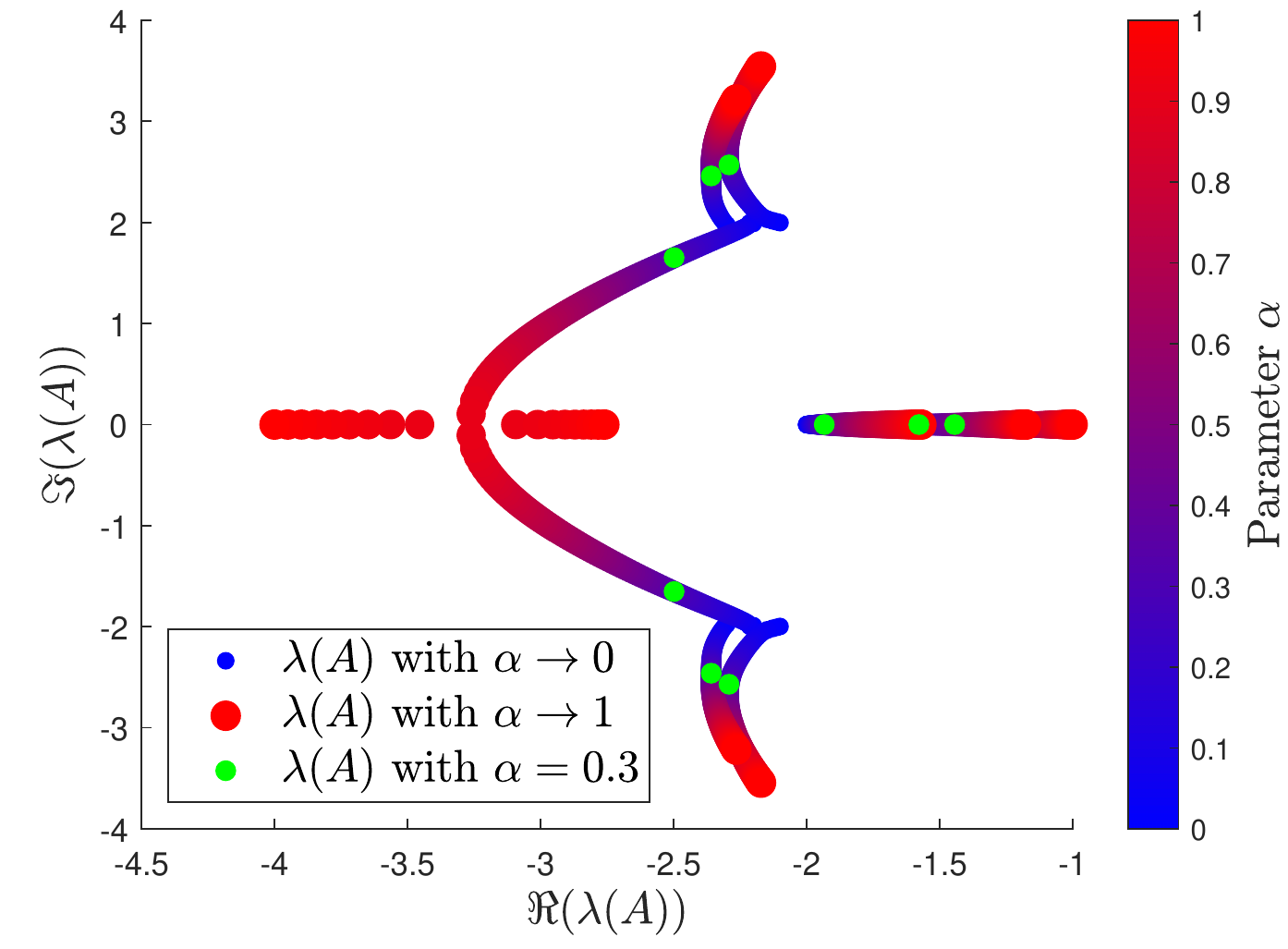}\\
    \includegraphics[width=\columnwidth]{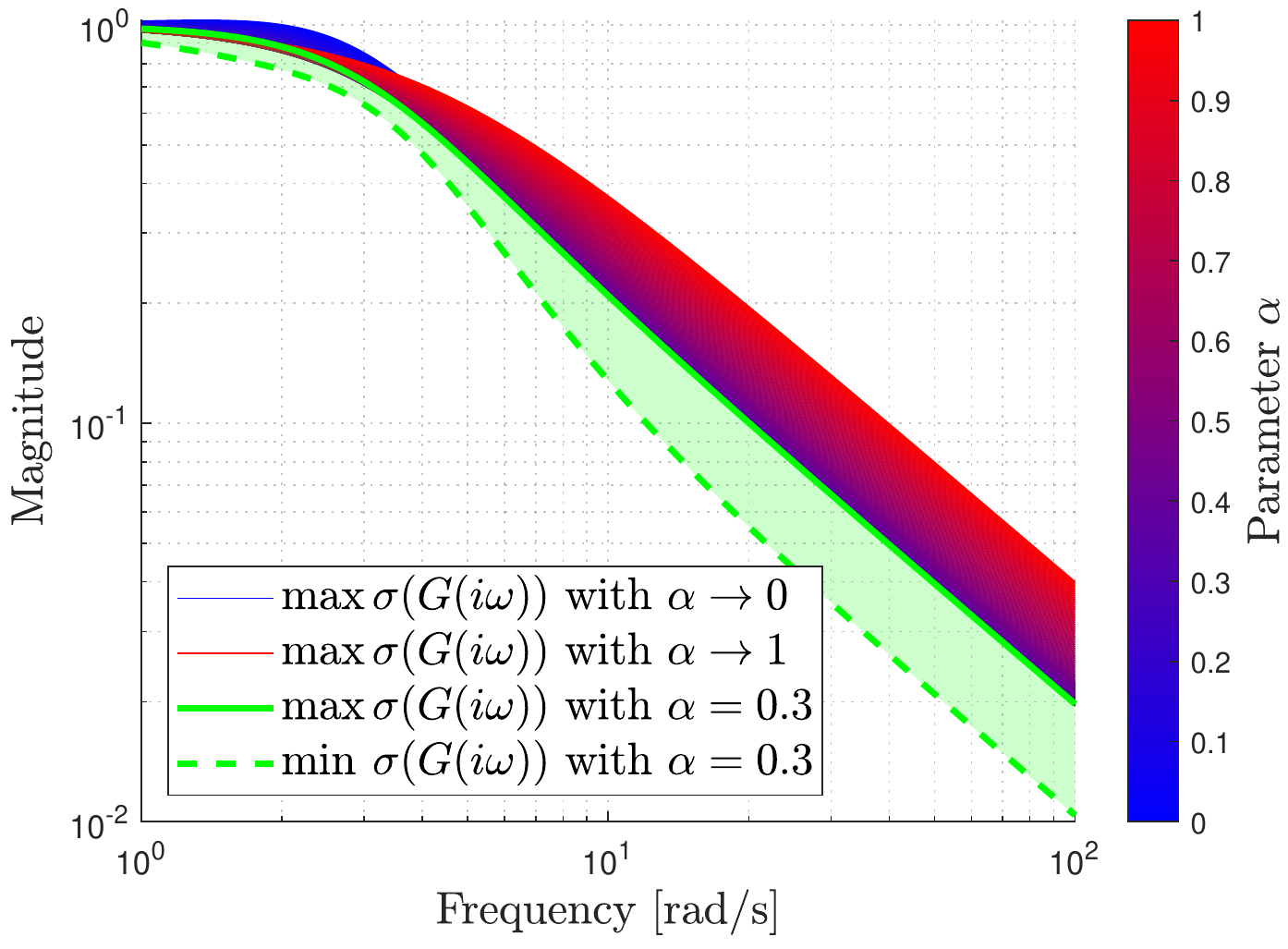}}
    \inArxiv{\includegraphics[width=0.5\textwidth]{figures/specplot.pdf}%
    \includegraphics[width=0.5\textwidth]{figures/sigmaplot.pdf}}
    \vspace{-15pt}
    
    \caption{\textit{Top:} Spectrum of the system matrix  $A$ governing the local error dynamics as a function of $\alpha$ for the nominal tuning in Sec.~\ref{sec:ideal}, with $\alpha=0.3$ marked in green. \textit{Bottom:} Singular values of the error dynamics from the inputs $\delta_0$ to $\tilde{x}$, with the area between the smallest and largest singular value at $\alpha=0.3$ in green.}
    \label{fig:tuning}
}
\inACC{  \begin{figure}[t!]\sigmaplot\end{figure}}
\inArxiv{\begin{figure}[h!]\sigmaplot\end{figure}}
In Fig.~\ref{fig:tuning}, we show how the spectrum of $A$, here denoted $\lambda(A)$, changes in the complex plane when varying the parameter $\alpha$ subject to the nominal tuning and realization in Sec.~\ref{sec:ideal} and a stationary rotation $R$. Note, that the error dynamics are time invariant if and only if $R$ is time invariant. We also show the maximum singular value from the gyroscopic noise input $\delta_0$ to the local observation errors $\tilde{x}$. That is, with the transfer function $G(s) = (sI-A)^{-1}B$, we compute the singular values $\sigma(G(i\omega)) = \sqrt{\lambda(G(-i\omega)^{\top}G(i\omega))}$ as a function of the frequency $\omega$.

The location of the poles of the linearized error dynamics behave highly non-trivially as a function of the observer parameters $\{k_a,k_b,k_R,k_l,\alpha,J\}$, and that when fixing the nominal parameters and varying $\alpha$, we get a relatively balanced system with real-parts of the spectrum ranging from -1.5 to -2.5 (as expected from the ULES property). Importantly, when looking at the influence of the gyroscopic noise on the observation errors, we note that noise with DC characteristics will still affect the observation errors, but that this noise is greatly suppressed for higher frequencies. It is also interesting to note that we should pick a lower $\alpha$ if the noise has significant spectral density at higher frequencies, and that it should be picked higher if the noise is of a DC nature. For this tuning, we found that an $\alpha=0.3$ yielded a good trade-off based on this (and several other) sigma plots. If using the estimator Proposition~\ref{prop:2} without filtering, we would have unit amplification across the entire spectrum, whereas low-pass filtering would suppress the noise after a cutoff frequency, but introduce a phase lag in the attitude rate estimate. This is completely avoided with the observer in Proposition~\ref{prop:4}, where we get the best of both worlds: perfect tracking under ideal conditions, and suppression of the high-frequent measurement noise. This analysis, applied to all of the parameters in turn and selecting combinations yielding an attenuation of the noise-to-state gains gave rise to the tuning in Sec.~\ref{sec:ideal}.

%
%

\section{Conclusions}\label{sec:conclusions}\label{sec:final}
In this paper, we first present an observer to estimate the angular momentum of the attitude dynamics without using measurements of angular velocities. We subsequently fuse this observer with a classical result of Mahony, generating an observer that is capable of estimating the attitude, attitude rate, and gyroscopic bias with UaGAS and ULES properties of the resulting error dynamics. Furthermore, we demonstrate that the combined observer has an edge over the two separate observers in terms of the asymptotic observer errors. Specifically, with the combined observer, we get good attenuation of high-frequent measurement noise, obtaining perfect tracking under ideal conditions, and having implicit robustness afforded by the uniform stability properties shown by the Matrosov result.

Importantly, this observer can be used to extend prior work on filtered output feedback in~\citep{lefebergreiffrobertsson2020} to a setting in which the gyroscopic biases are estimated and accounted for. This will be done in our future work.

\if\useieeelayout1
\begin{ack}
We thank Thor Inge Fossen for inspiring this paper.
\end{ack}
\else
\section{Acknowledgement}
We thank Thor Inge Fossen for inspiring this paper during his visit to Lund University.
\fi

\bibliography{references}

\appendix

\newcommand\defeq[1]{\underset{\eqref{#1}}{=}}
\newcommand\mineig[1]{\ubar{\lambda}(#1)}
\newcommand\maxeig[1]{\bar{\lambda}(#1)}

\section{Supplementary details for Proof~\ref{pf:2}}\label{app:A} 
In this section, we give supplementary details for Proof~\ref{pf:2}, defining the constants of the proof as a function of the known parameters and assumed bounds. The maximum and minimum eigenvalues of a real symmetric matrix $J$ are denoted $\maxeig{J}$ and $\mineig{J}$, respectively. Further, $M=\sum_{i=1}^nk_iv_iv_i^\top$ with distinct eigenvalues $\lambda_i$, i.e., $\lambda_3>\lambda_2>\lambda_1>0$. Let  $D=\diag(r_2\lambda_2+r_3\lambda_3,r_1\lambda_1+r_3\lambda_3,r_1\lambda_1+r_2\lambda_2)$ with appropriate $r_i\in\{-1,1\}$, and take $\bar D = R^{\top}DR$.

\subsection{The inequality in~\eqref{eq:p2firstbound}}
We start by showing the first inequality in the context of Proof~\ref{pf:2}. Here: $\tilde{r}_k$ is bounded by definition; $\tilde{\ell}$ is bounded for all times as the Lyapunov function is negative semi-definite along the solutions of the error dynamics; the attitude rates $\omega$ and accelerations $\dot\omega$ are both bounded by assumption. In summary, $\exists K_i>0$ for $i=1,...,4$, such that
\begin{align*}
    \|\tilde{r}_k\|_2&\leq \Big(\sum_{i=1}^n k_i^2\Big)^{1/2}\triangleq  K_1,&
    \|\tilde{\ell}\|_2&\leq \Big(2V_1(0)-k_\ell\sum_{i=1}^n{k_i}\|\tilde Rv_i-v_i\|_2^2\Big)^{1/2}\triangleq  K_2, &
    \|\omega\|_2&\leq K_3,&
    \|\dot \omega\|_2&\leq K_4.
\end{align*}
Thus, $\dot{\hat\ell}$ is bounded,
\begin{align*}
    \|\dot{\hat\ell}\|_2&\leq \|J\dot{\omega}\|_2+\|S(J\omega)\omega)\|_2+k_\ell \|J^{-1}\tilde r_k\|_2\leq \maxeig{J}K_4 + \maxeig{J}K_3^2+k_\ell \mineig{J}^{-1}K_1 \triangleq K_5
\end{align*}
Let $\tilde r_k = R^{\top} r_k$, then
\begin{align}
\tilde r_k&=
\sum_{i=1}^n k_iS(\hat R^\top v_i)R^\top v_i
\defeq{eq:rotid}
R^{\top}\sum_{i=1}^n k_iS(R\hat R^\top v_i)v_i
\defeq{eq:error}
R^{\top}\sum_{i=1}^n k_iS(\tilde R^\top v_i)v_i=R^{\top}r_k.
\end{align}
In light of Lemma~\ref{lem:nul}, $\|D\|_2=\|\bar D\|_2=K_6$ and this constant is known in the observer tuning $\{(k_i,v_i)\}_{i=1}^N$. Now,
\begin{align}
\dot{r}_k &=\sum_{i=1}^n k_iS(\dot{\tilde R}^\top v_i)v_i\defeq{eq:rkdot} - 
DR(J^{-1}R^\top \tilde\ell-k_R\tilde r_k)
\qquad\Rightarrow \qquad
\|\dot{r}_k\|_2  \leq K_6(\mineig{J}^{-1}K_2+k_RK_1)\triangleq K_7.
\end{align}
By the chain rule
\begin{align}\label{eq:tilderdotapp}
\dot{\tilde r}_k&=
S(\omega)^{\top}{R}^{\top}r_k
-R^{\top}\dot r_k\quad\Rightarrow \quad
    \|\dot{\tilde r}_k\|_2 \leq K_3K_1+K_7\triangleq K_8.
\end{align}
Differentiating~\eqref{eq:tilderdotapp}, once more
\begin{align}
\ddot{\tilde r}_k  
&=
(S(\omega)^2+S(\dot\omega)^{\top}){R}^{\top}r_k 
+S(\omega)^{\top}{R}^{\top}\dot r_k\\
&\qquad+[S(\omega)^{\top}\bar D+\bar D S(\omega)](J^{-1}R^\top \tilde\ell-k_R\tilde r_k)
+\bar D (J^{-1}S(\omega)^{\top}R^\top \tilde\ell-k_R\dot{\tilde r}_k)\quad\Rightarrow\notag
\\
\|\ddot{\tilde r}_k\|_2& \leq 
K_1(K_3^2+K_4)
+ K_3K_7
+ 2K_3K_6(\maxeig{J^{-1}}K_2+k_RK_1)
+ K_6(\maxeig{J^{-1}}K_2K_3+k_RK_7)
 \triangleq K.
\end{align}
This is the constant $K>0$ that appears in~\eqref{eq:p2firstbound}, which is computable in the initial errors and tuning parameters.

\subsection{The inequality in~\eqref{eq:p2secondbound}}
The second inequality follows directly from~\eqref{eq:tilderdotapp}. We obtain
\begin{subequations}
\begin{align}
    -\|\dot{\tilde r}_k\|_2^2 &= -r_k^{\top} RS(\omega)S(\omega){R}^{\top}r_k + 2r_k^{\top}R^{\top}\dot r_k - \|\dot r_k\|^2_2\\
    &\leq  K_3^2\|\tilde r_k\|^2_2 + 2K_8\|\tilde r_k\|_2 - \|DR(J^{-1}R^\top \tilde\ell-k_R\tilde r_k)\|^2_2\\
    &\leq  K_3^2\|\tilde r_k\|^2_2 + 2K_8\|\tilde r_k\|_2
    - \tilde\ell^{\top} R J^{-1} R ^{\top}D^2RJ^{-1}R^\top \tilde\ell
    +2k_RK_2 \maxeig{J^{-1}} K_6\|\tilde r_k\|_2
    +k_R^2\|\tilde r_k\|^2_2\\
    &\triangleq 
    - \gamma\|\tilde\ell\|_2^2 + K_{9}\|\tilde r_k\|_2 
    \label{eq:52d}
\end{align}
\end{subequations}
thus
\begin{align}
    -\|\dot{\tilde r}_k\|_2^2+K\|\tilde r_k\|_2 & \leq 
    - \gamma\|\tilde\ell\|_2^2 + K_{9}\|\tilde r_k\|_2 
    +K\|\tilde r_k\|_2 \triangleq 
    -\gamma \|\tilde\ell\|^2_2+\bar K\|\tilde r_k\|_2,
\end{align}
where
\begin{align}
\gamma &=\mathrm{inf}_{R\in SO(3)} \mineig{RJ^{-1}R^\top D^2RJ^{-1}R^{\top}},&
\bar K &= K_3^2K_1+2K_8+2k_RK_2 \maxeig{J^{-1}} K_6+k_R^2K_1,
\end{align}
are the constants in~\eqref{eq:p2secondbound}. Note that $D$ may be indefinite, but $D^2$ is positive definite, thus $\gamma>0$ is a positive constant.

\subsection{Proof of ULES}
The proof of ULES uses the main ideas in~\citep{wulee2016} and consists of three main steps:
\begin{itemize}
\item[(i)] Show that for sufficiently small $\bar \epsilon>0$, there exists positive constants $c_1,c_2$ for which
\begin{align}
c_1\|\tilde r_k\|_2^2\leq & k_\ell\sum_{i=1}^n\myfrac{k_i}{2}\|\tilde Rv_i-v_i\|_2^2\leq c_2\|\tilde r_k\|_2^2, \qquad \forall \|\tilde r_k\|_2\leq \bar\epsilon.
\end{align}
\item[(ii)] As $\dot{V}_1\leq 0$, let $\tilde\omega = J^{-1}R^\top \tilde\ell$, and that $\exists\;\epsilon_1>0$ defining a set $\mathcal{S}(\epsilon_1) = \{(\tilde r_k, \tilde \omega)\in\mathbb{R}^6\;|\;V_1\leq \epsilon_1\}$ for which
\begin{equation}\label{eq:eps1}
\sup_{(\tilde r_k,\tilde \omega)\in\mathcal{S}(\epsilon_1)} \|\tilde r_k\|_2\leq \bar \epsilon.
\end{equation}
\item[(iii)] Finally, define $z^{\top}=(\tilde r_k^{\top}, \tilde \omega^{\top})\in\mathbb{R}^6$ and consider a composite function $V_3 = V_1 + \epsilon V_2$ for some small $\epsilon > 0$. Show that for sufficiently small $\epsilon$, there exists positive definite matrices $M_1,M_2,W$, such that
\begin{align}
z^{\top}M_1z\leq &V_3 \leq z^{\top}M_2z, &
&\dot V_3 \leq -z^{\top}Wz, & \forall& z \in \mathcal{S}(\epsilon).
\end{align}
ULES of the origin $z=0$ then follows by application of {\citet[Th.~4.10]{Khalil}}.
\end{itemize}

\subsubsection{Details on step (iii)}
The first two steps are straight-forward, but some additional details are provided for step (iii). Recall that
\begin{align}\label{eq:app:rtildedot}
\dot{\tilde r}_k&=
S(\omega)^{\top}\tilde r_k
-R^{\top}\dot r_k = 
S(\omega)^{\top}\tilde r_k
+\bar D(\tilde\omega-k_R\tilde r_k),
\end{align}
thus
\begin{align}
V_2&=-\tilde r_k^\top[S(\omega)^{\top}\tilde r_k
+\bar D(\tilde\omega-k_R\tilde r_k)]=
-\tilde r_k^\top \bar D\tilde\omega+ k_R\tilde r_k^\top \bar D\tilde r_k,
\end{align}
and we obtain
\begin{equation}
z^{\top}
\underbrace{\begin{pmatrix}
\begin{bmatrix}
c_1I & 0 \\ 0 & J^2
\end{bmatrix}+\begin{bmatrix}
\epsilon k_R\bar D & -\tfrac12\epsilon\bar D\\ -\tfrac12\epsilon \bar D & 0
\end{bmatrix}
\end{pmatrix}}_{\triangleq M_1}z \leq V_3\leq z^{\top}\underbrace{\begin{pmatrix}
\begin{bmatrix}
c_2I & 0 \\ 0 & J^2
\end{bmatrix}+\begin{bmatrix}
\epsilon k_R\bar D & -\tfrac12\epsilon\bar D\\ -\tfrac12\epsilon \bar D & 0
\end{bmatrix}
\end{pmatrix}}_{\triangleq M_2}z.
\end{equation}
Sufficient conditions for $M_i>0$ can be expressed as a bound on $\epsilon$ in $\{c_1,c_2,k_R,D,J\}$. Taking the Schur complement
\begin{align}
\hspace{-8pt}M_i>0&\;\Leftrightarrow\; c_iI+\epsilon k_R\bar D>0\;\land\; J^2 - \tfrac14 \epsilon \bar D (c_iI+\epsilon k_R\bar D)^{-1}\epsilon \bar D>0
\;\Leftarrow\;  4\mineig{J^2}(c_iI+\epsilon k_R D)D^{-2} - \epsilon^2  >0,
\end{align}
as $D$ is a diagonal matrix and all constants are positive (definite). If we let $D=\mathrm{diag}(d_1,d_2,d_3)$, the condition is
\begin{equation}\label{eq:eps2}
\epsilon < \epsilon_2\triangleq \min_{\substack{i\in\{1,2\}\\j\in\{1,2,3\}}}\begin{Bmatrix}\dfrac{2\mineig{J^2} k_R}{d_j}\begin{pmatrix}1 + \sqrt{1 + \dfrac{c_i}{\mineig{J^2} k_R^2}}\end{pmatrix}\end{Bmatrix}.
\end{equation}

To find the quadratic form bounding $\dot{V}_3$, we first consider the terms of $\dot{V}_2=-\| \dot{\tilde r}_k\|_2^2-\tilde r_k^\top \ddot{\tilde r}_k$. It can be shown that
\begin{align}
 \|\dot{\tilde r}_k\|_2^2
&=
\begin{bmatrix}
\tilde r_k\\
\tilde \omega
\end{bmatrix}^{\top}
\begin{bmatrix}
k_R^2I-S(\omega)^2 & (S(\omega)-2k_R I)\bar D \\
\bar D(S(\omega)^{\top}-2k_RI) & \bar D^2
\end{bmatrix}
\begin{bmatrix}
\tilde r_k\\
\tilde \omega
\end{bmatrix}.
\end{align}
Furthermore, taking the time-derivative of~\eqref{eq:app:rtildedot}, we get
\begin{align}
\ddot{\tilde r}_k&=
S(\dot\omega)^{\top}\tilde r_k + S(\omega)^{\top}\dot{\tilde r}_k
+\dot{\bar D}(\tilde\omega-k_R\tilde r_k)
+\bar D(\dot{\tilde\omega}-k_R\dot{\tilde r}_k),
\end{align}
and
\begin{align}
 \dot{\tilde\omega}&
=J^{-1}S(\omega)^{\top}J\tilde\omega - k_\ell J^{-2}\tilde r_k.
\end{align}
With these expressions, one can obtain
\begin{align}
\tilde r_k^{\top}\ddot{\tilde r}_k&=
\begin{bmatrix}
\tilde r_k\\
\tilde \omega
\end{bmatrix}^{\top}
\begin{bmatrix}
S(\omega)^2 + \bar W_{11} & \bar W_{12} \\
\star & 0
\end{bmatrix}
\begin{bmatrix}
\tilde r_k\\
\tilde \omega
\end{bmatrix},
\end{align}
where bounds on $\|{\bar W}_{11}\|_2$ and $\|{\bar W}_{12}\|_2$ can be expressed $J$, $D$, and the assumed bound of $\omega$. 
Now, we have that
\begin{align}
\tilde r_k^{\top}\ddot{\tilde r}_k + \|\dot{\tilde r}_k\|_2^2=\begin{bmatrix}
\tilde r_k\\
\tilde \omega
\end{bmatrix}^{\top}
\begin{bmatrix}
k_R^2I+\bar W_{11} & (S(\omega)-2k_R I)\bar D + \bar W_{12}\\
\star  & \bar D^2
\end{bmatrix}
\begin{bmatrix}
\tilde r_k\\
\tilde \omega
\end{bmatrix}\triangleq
\begin{bmatrix}
\tilde r_k\\
\tilde \omega
\end{bmatrix}^{\top}
\begin{bmatrix}
\bar{\bar W}_{11} & \bar{\bar W}_{12} \\
\star  & \bar D^2
\end{bmatrix}
\begin{bmatrix}
\tilde r_k\\
\tilde \omega
\end{bmatrix},
\end{align}
where similarly $\|\bar{\bar W}_{11}\|_2$ and $\|\bar{\bar W}_{12}\|_2$ are bounded. Subsequently
\begin{align}
\dot V_3 &= \dot V_1 + \epsilon \dot V_2 = -k_\ell k_R\|\tilde r_k\|_2^2 - \epsilon(\tilde r_k^{\top}\ddot{\tilde r}_k + \|\dot{\tilde r}_k\|_2^2) 
=-\begin{bmatrix}
\tilde r_k\\
\tilde \omega
\end{bmatrix}^{\top}\underbrace{\epsilon
\begin{bmatrix}
\epsilon^{-1}k_\ell k_R I +\bar{\bar W}_{11} & \bar{\bar W}_{12} \\
\star  & \bar D^2
\end{bmatrix}}_{\triangleq { W}}
\begin{bmatrix}
\tilde r_k\\
\tilde \omega
\end{bmatrix}.
\end{align}
As such, we get a conservative but sufficient condition for $W>0$ by picking a sufficiently small $\epsilon>0$. Specifically,
\begin{equation}\label{eq:eps3}
W>0\Leftrightarrow \bar D^2>0\;\;\land \;\;\epsilon^{-1}k_\ell k_R I +\bar{\bar W}_{11} - \bar{\bar W}_{12}^{\top}\bar D^{-2} \bar{\bar W}_{12} > 0\Leftarrow \epsilon< \epsilon_3\triangleq k_\ell k_R(\|\bar{\bar W}_{12}^{\top}\|_2\|\bar D^{-2}\|_2 \|\bar{\bar W}_{12}\|_2 + \|\bar{\bar W}_{11}\|_2)^{-1}.
\end{equation}
Using {\citep[Th.~4.10]{Khalil}} on $\mathcal{S}(\epsilon)$ with any $\epsilon < \min\{\epsilon_1,\epsilon_2, \epsilon_3\}$ from~\eqref{eq:eps1},~\eqref{eq:eps2} and~\eqref{eq:eps3} completes the proof.

\section{Parameter Distributions used in the Monte Carlo Simulations}\label{sec:parameters}
In this section, we let $\mathcal{N}(x; \mu, \Sigma)$ denote a Gaussian probability density function (PDF) in $x$ with mean $\mu\in\mathbb{R}^n$ and covariance $\Sigma\in\mathbb{R}^{n\times n}$. We let $\mathcal{U}(x; I)$ be a uniform PDF in $x$ that samples every element of a closed interval $I\subset \mathbb{R}^n$, uniformly and independently in each dimension. In practice, we accomplish this for SO(3) by drawing an un-normalized quaternion $\mathcal{N}(q; 0,I)$, normalizing this, and embedding it in SO(3). We refer to this as $\mathcal{U}(R; \textup{SO(3)})$.

The parameters $\theta\hspace{-2.5pt}=\hspace{-3.5pt} \{R_0,\omega_0,b,\hat{R}_0,\hat{\omega}_0,\hat{b}_0\}$ are sampled from \inArxiv{a distribution with probability density function}
\begin{align*}
    p(\theta) =\;& \mathcal{U}(R_0; \textup{SO(3)})\;\mathcal{N}(b_0; 0, I)\;\mathcal{N}(\omega_0; 0, 0.1I)\inACC{\times}\ifacbreak \mathcal{U}(\hat{R}_0; \textup{SO(3)})\;\mathcal{N}(\hat{b}_0; 0, I)\;\mathcal{N}(\hat{\ell}_0; 0, I),
\end{align*}
the inertia $J$ is constructed by sampling a random symmetric positive semi-definite matrix $J_A$ with spectrum $\{\lambda_1,\lambda_2,\lambda_3\}$, where $0=\lambda_1\leq \lambda_2\leq \lambda_3=1$, and letting $J = 0.5(J_A + I)$. We let $v_1 = (0,0,-1)^{\top}$, sample $\mathcal{N}(\bar{v}_2; 0, I)$, setting its last element to -0.1, and normalizing it, such that $v_2 = \bar{v}_2/\|\bar{v}_2\|_1$. We then take $v_3=v_1\times v_2$.

In the ideal setting (Sec.~\ref{sec:ideal}), the outputs in~\eqref{eq:output} are sampled continuously without noise, and we run the simulation with a fixed-point RK4 solver over $t\in[0,10]$ seconds.

In the Monte Carlo simulations, we add noise terms $n_i$, as
\begin{align*}
y_0(hk)&=\omega(hk)+b(hk)+n_0(hk) \\ \bar{y}_i(hk)&=R(hk)^\top v_i+n_i(hk) & i=1,\dots, n,\\ {y}_i(hk)&=\bar{y}_i(hk)/\|\bar{y}_i(hk)\|_2 & i=1,\dots, n,
\end{align*}
and we take this noise to be zero-mean Gaussian distributed, uncorrelated, sampled from $\mathcal{N}(n_i(hk); 0, 0.01I)$. We sample these outputs at a rate of 500 Hz (i.e. $h=0.002$ s), but run the observer prediction at a rate of 1 kHz.

\section{An Equivalent Discrete-Time Quaternion Formulation}\label{app:B}

Just as in~\protect{\cite[Appendix B]{mahony2008nonlinear}}, it is straightforward to give an equivalent representation of the filters when integrating the attitude as a quaternion. The set of quaternions is $\mathbb{H}=\{q = (q_w, q_{v})\in\mathbb{R}\times \mathbb{R}^3:|q| = 1\}$, and we use the Hamilton construction with quaternion representing a right-handed rotation (see, e.g.,~\citep{greiff2021nonlinear}). The group $\mathbb{H}$ is 2-to-1 homomorphic to $\textup{SO(3)}$,
\begin{equation*}
E:\mathbb{H}\mapsto \textup{SO(3)}, \quad E(q)=(q_w^2-q_v^{\top}q_v)I + 2q_vq_v^{\top} + 2q_wS(q_v).
\end{equation*}
The attitude kinematics of the quaternion, i.e., the differential equation preserving $q(t+\delta)\in\mathbb{H}$ for $\delta\geq 0$ is
\begin{equation*}
    \dot{q} = \frac{1}{2}Q(q)\begin{bmatrix}
             0\\ \omega
    \end{bmatrix}, \quad Q(q) = Iq_w +  \begin{bmatrix}
             q_w & -q_v^{\top}\\q_v & S(q_v)
    \end{bmatrix}.
\end{equation*}
As such, to implement an observer with an quaternion attitude representation $\hat{q}(t)$ such that $\hat{R}(t)=E(\hat{q}(t))$, we only need to replace~\eqref{eq:dotRprop4} in Proposition~\ref{prop:4} by
\begin{equation*}
    \dot{\hat{q}} = \frac{1}{2}Q(\hat{q})\begin{bmatrix}
             0\\
             \alpha J^{-1}\tilde{\delta}_L+y_0-\hat{b}-k_R\tilde{r}_k
    \end{bmatrix},
\end{equation*}
and modify the computation of $\tilde{r}_k$ in~\eqref{eq:tilderk} as
\begin{align*}
\tilde r_k=\sum_{i=1}^n k_iS(E(\hat{q})^\top v_i)y_i.
\end{align*}

In any practical implementation, the observer update would need to be discretized. Here, a sufficiently slow update rate with a sufficiently simple discretization will lead to numerical artifacts that may become a dominating factor in the noise floor of the error dynamics. Instead of a forward Euler scheme, as is commonly used in practice, we suggest the use of an RK scheme of higher order with a projection onto $\mathbb{H}$ on each time step, or a Crouch-Grossman integrator which performs the integration directly on $\mathbb{H}$ (refer to the discussion in~\protect{\cite[Chapter 2.4]{greiff2021nonlinear}}).

\newcommand*{\fvtextcolor}[2]{\textcolor{#1}{#2}}
To simplify implementations of the theoretical results, we include the observer as Matlab code with a fixed-step RK4 integrator. Proposition~\ref{prop:4} can be implemented as:
\if\useieeelayout1
\begin{Verbatim}[commandchars=&\{\}]
&fvtextcolor{blue}{function} Xkp1 = obs_ODE(Xk,y0,y1,y2,y3,tau)
    &fvtextcolor{green!50!black}{% Define observer parameters}
    v1 = ; v2 = ; v3 = ; k1 = ; k2 = ; k3 = ;
    J = ; kr = ; kl = ; ka = ; kb = ; alpha = ;
    &fvtextcolor{green!50!black}{% Required functions}
    S = @(u) [    0,-u(3), u(2);...
               u(3),    0,-u(1);...
              -u(2), u(1),   0];
    Q = @(q) eye(4).*q(1) + ...
         [q(1),-q(2:4)'; q(2:4),  S(q(2:4))];
    E = @(q) (q(1)^2-q(2:4)'*q(2:4))*eye(3)+...
         2*q(2:4)*q(2:4)'+2*q(1)*S(q(2:4));
    &fvtextcolor{green!50!black}{% Process arguments}
    R = (k1*v1*v1' + k2*v2*v2' + k3*v3*v3')\...
        (k1*v1*y1' + k2*v2*y2' + k3*v3*y3');
    bhat = Xk(1:3);
    lhat = Xk(4:6);
    qhat = Xk(7:10);
    &fvtextcolor{green!50!black}{% Observer update}
    rtilde = k1*S(E(qhat)'*v1)*y1 +...
             k2*S(E(qhat)'*v2)*y2 +...
             k3*S(E(qhat)'*v3)*y3;
    deltaL = R'*lhat - J*(y0 - bhat);
    deltaR = alpha * inv(J) * deltaL +...
             y0 - bhat - kr*rtilde;
    bhatdot = kb*rtilde -...
              alpha * kb * ka * J * deltaL;
    lhatdot = R * (tau - kl*  inv(J) * rtilde...
              -(1-alpha) * kl * ka * deltaL);
    qhatdot = Q(qhat) * [0; deltaR/2];
    Xkp1    = [bhatdot; lhatdot; qhatdot];
&fvtextcolor{blue}{end}
\end{Verbatim}
\else
\begin{Verbatim}[commandchars=&\{\}]
&fvtextcolor{blue}{function} Xkp1 = obs_ODE(Xk,y0,y1,y2,y3,tau)
    &fvtextcolor{green!50!black}{% Define observer parameters}
    v1 = ; v2 = ; v3 = ; k1 = ; k2 = ; k3 = ;
    J = ; kr = ; kl = ; ka = ; kb = ; alpha = ;
    &fvtextcolor{green!50!black}{% Required functions}
    S = @(u) [    0,-u(3), u(2); u(3),    0,-u(1);  -u(2), u(1),   0];
    Q = @(q) eye(4).*q(1) + [q(1),-q(2:4)'; q(2:4),  S(q(2:4))];
    E = @(q) (q(1)^2-q(2:4)'*q(2:4))*eye(3)+2*q(2:4)*q(2:4)'+2*q(1)*S(q(2:4));
    &fvtextcolor{green!50!black}{% Process arguments}
    R = (k1*v1*v1' + k2*v2*v2' + k3*v3*v3') \ (k1*v1*y1' + k2*v2*y2' + k3*v3*y3');
    bhat = Xk(1:3);
    lhat = Xk(4:6);
    qhat = Xk(7:10);
    &fvtextcolor{green!50!black}{% Observer update}
    rtilde = k1*S(E(qhat)'*v1)*y1 + k2*S(E(qhat)'*v2)*y2 + k3*S(E(qhat)'*v3)*y3;
    deltaL = R'*lhat - J*(y0 - bhat);
    deltaR = alpha * inv(J) * deltaL + y0 - bhat - kr*rtilde;
    bhatdot = kb*rtilde - alpha * kb * ka * J * deltaL;
    lhatdot = R * (tau - kl*  inv(J) * rtilde -(1-alpha) * kl * ka * deltaL);
    qhatdot = Q(qhat) * [0; deltaR/2];
    Xkp1    = [bhatdot; lhatdot; qhatdot];
&fvtextcolor{blue}{end}
\end{Verbatim}
\fi
The observer update with a fixed-step RK4 scheme is then:
\begin{Verbatim}[commandchars=&\{\}]
&fvtextcolor{blue}{function} Xkp1 = update(Xk,h,y0,y1,y2,y3,tau)
    &fvtextcolor{green!50!black}{% 4th order Runge-Kutta update}
    k1 = obs_ODE(Xk         ,y0,y1,y2,y3,tau);
    k2 = obs_ODE(Xk + h/2*k1,y0,y1,y2,y3,tau);
    k3 = obs_ODE(Xk + h/2*k2,y0,y1,y2,y3,tau);
    k4 = obs_ODE(Xk + h/1*k3,y0,y1,y2,y3,tau);
    Xkp1 = Xk + h/6 * (k1 + 2*k2 + 2*k3 + k4);
    &fvtextcolor{green!50!black}{% Projection to a unit quaternion}
    Xkp1(7:10) = Xkp1(7:10) /norm(Xkp1(7:10));
&fvtextcolor{blue}{end}
\end{Verbatim}
Here, we include a projection onto $\mathbb{H}$ which becomes necessary when tuning the observer with higher gains. For all of the simulations, the attitude was integrated directly on $\textup{SO(3)}$, but the RK-method above produces identical results and is well suited for practical implementations.

\end{document}